\documentclass[12pt]{article}
\newcommand{\beq}{\begin{equation}}
\newcommand{\eeq}{\end{equation}}
\newcommand{\ej}{\cal E}
\newcommand{\hej}{\hat {\ej}}
\begin{document}

\title{\bf On form-factors in Sin(h)-Gordon theory.}
\author{A.Rosly and K.Selivanov}

\date{ITEP-TH-77/97}
\maketitle

\begin{abstract}
We present here an explicit classical solution of the type of perturbiner 
in Sin(h)-Gordon model. This solution is a generating function for
form-factors in the tree approximation.
\end{abstract}

Perturbiner is a solution of classical field equations which is a
generating function for tree form-factors -- that is objects of the
type of $< 1, ..., N|\phi(x)|0>$ -- in the theory. A convenient
definition of this solution -- formally independent of considering
Feynman diagrams -- was given in \cite{RS1}.
Verbally, it sounds as follows.
Take a solution of the free (linear) field equation in the theory under consideration 
in the form of superposition of a set of plane waves with every plane wave
multiplied by a nilpotent coefficient. The corresponding perturbiner is a complex 
solution of the full (nonlinear) field equation which is 
polynomial in the (nilpotent) plane waves, first order term of the polynomial being 
the solution of the free field equation. The plane waves are nothing but the asymptotic 
wave functions of the one-particle states included in the form-factors which the
perturbiner is the generating function for. The nilpotency assumes that the perturbiner
 is the generating 
function for form-factors without identical one-particle states. Obviously, there is no 
loss of generality in the assumption of nilpotency.
Of course, it is very well known that tree amplitudes (form-factors)
can in principle be obtained from classical field equations, see, for
instance, \cite{FS}, \cite{IZ}, where the appropriate solution was
defined by the Feynman-type boundary conditions. Our definition is
different and has an advantage that an infinite-dimensional
space of space-time functions is substituted by the
finite-dimensional space of polynomials in a finite number of
nilpotent variables. In particular, the existence and uniqueness
are almost automatic in our case.

Of course, in generic theory one cannot go further than the
uniqueness and existence theorem and the common perturbation
procedure. Nevertheless, the perturbiner happens to be very easy to
construct in cases when the field equations are integrable, that is,
essentially, when there is a zero-curvature representation. Thus we
have constructed perturbiner for the self-duality equations in
Yang-Mills theory \cite{RS2}, in gravity \cite{RS3} and in Yang-Mills
theory interacting with gravity \cite{S1}. The zero-curvature
representation in the above cases is provided by the twistor
construction \cite{Penrose}, \cite{Ward}, \cite{AHS}. 
 The way of reducing from the
generic perturbiner to the self-dual one is obvious: one includes into the
plane wave solutions of the linearized field equations only 
self-dual plane waves, which
is equivalent to describing only amplitudes with on-shell particles of 
a given helicity (say, of the positive one). The particles of the opposite helicity 
can then be
added one-by-one perturbatively (see \cite{RS2} and \cite{S2}).
{\footnote{In the case of Yang-Mills theory, the idea to use the 
self-duality equations to describe the so-called like-helicity amplitudes
was first formulated in \cite{Ba} and, independently, in \cite{Se}.
 In \cite{Ba} it was basically shown that  the self-duality
equations reproduce the recursion relations for tree like-helicity gluonic 
form-factors (also called ``currents''), obtained originally 
in ref.\cite{BG} from the Feynman diagrams; the  corresponding solution of 
self-duality equations was than obtained in terms of the known solutions 
refs.\cite{BG} of the recursion relations 
for the ``currents''. In ref.\cite{Se} an example of self-dual perturbiner was
obtained in the SU(2) case by a 'tHooft ansatz upon further restriction on the
on-shell particles included. A bit later  a similar solution was studied
in ref.\cite{KO}, where consideration was based
on solving  recursion relations analogous to refs. \cite{BG}.}}

In this letter we describe construction of the perturbiner in the
Sin(h)-Gordon model (apparently, since perturbiner is a complex
solution, there is no difference between Sin-Gordon and Sinh-Gordon 
cases). Thus
we directly obtain the generating function for the tree form-factors
in the theory. Although, the form-factors in the model are 
known exactly \cite{F.S1}, not only in the tree approximation, we
still think that our construction makes some sense. Firstly, we could
not find in the literature a closed explicit expressions for the tree
limit of the form-factors, so Eq.(\ref{24}) below fills this
gap. Say, in \cite{FMS} the exact expressions for the form-factors include
some polynomials for which only a recursion relation without explicit
general solution is given. And it seems not to be completely trivial
to extract tree factors from general formulae in \cite{F.S1}.
 Notice that what is called the classical limit in
\cite{F.S2} is different from the tree approximation. Secondly, the
popular approach to the exact form-factors is based on a set of
axioms \cite{F.S1}. Upon obtaining an exact solution for this set of
axioms it becomes a separate task to identify the corresponding
operator or even the corresponding field theory while our approach
starts from the field equations and is absolutely straightforward.
It is yet only a tree approximation, but notice, that the tree
contribution always serves as a skeleton for the complete quantum
expressions and sometimes factorizes out. For example, in \cite{Vergeles} 
the exact (with all quantum corrections) formulae for the
Sin(h)-Gordon form-factors are given in such a  form 
with the factorized (not given explicitly) tree contribution so that our results complement
the results of \cite{Vergeles}. In \cite{BFKL} the Yang-Mills amplitudes
in the so-called multi-Regge asymptotics are also given
in such a  factorized form with explicit tree factor. Thirdly, the present construction of the 
perturbiner in Sin(h)-Gordon theory is analogous to the construction of the
self-dual perturbiner in Yang-Mills theory and in gravity, in particular all
these constructions are based on the same universal solution for a sort of 
Riemann-Hilbert problem (see the last paragraph in Section 2). This, hopefully,
will open a new field of application for the exact 2d results.  Finally,
solutions of the type of perturbiner form another class of physically
motivated solutions, in addition to the solitonic solutions and to the finite-gap
solutions. Probably, they deserve study on their own right.

 The rest of this paper is organized as follows. In Section 1 we formulate
what type of solutions for the Sin(h)-Gordon equation we are looking for.
In Section 2 we lift this formulation
to the zero-curvature problem associated with the Sin(h)-Gordon equation.
In Section 3 we present solution of the problem. 

We would like to point out that our construction generalizes straightforwardly
to the 2d Toda theories.

\section{Definition}

Let us  consider the field equation
\beq
\label{1}
{\partial}{\bar{\partial}}{\phi}+
\frac{m^{2}}{\beta}{\rm sinh}{\beta}{\phi}=0
\eeq
where ${\partial}=\frac{\partial}{\partial z}$, 
${\bar{\partial}}=\frac{\partial}{\partial {\bar z}}$, and $z$, ${\bar z}$ are two 
coordinates. In what follows we put $m^{2}=1$ since $m^{2}$-dependence
can easily be restored. To define the perturbiner one  picks up a solution
\beq
\label{2}
{\phi}^{(1)}=\sum_{j=1}^{N} a_{j}e^{ik_{j}z+i\frac{1}{k_{j}}{\bar z}}=
\sum_{j} {\ej}_{j}
\eeq
of the free field equation 
\beq
{\partial}{\bar{\partial}}{\phi}^{(1)}+{\phi}^{(1)}=0
\eeq
The coefficients $a_{j}$ are assumed to be  commuting and nilpotent
formal variables, $a_{j}a_{l}=a_{l}a_{j}$, $a_{j}^{2}=0$. Index $j=1, \ldots , N$
numbers the one-particle states in the given set. 

The perturbiner ${\phi}^{ptb}$ is a (complex) solution of Eq.(\ref{1}) which is polynomial
in the variables ${\ej}_{j}$ entering Eq.(\ref{2}), first order term of the polynomial
being just  ${\phi}^{(1)}$ Eq.(\ref{2}). One can see that when the set of momenta
$k_{j}$ is generic this solution exists and is unique.

${\phi}^{ptb}(z, {\bar z}, \{k\}, \{a\})$ thus defined is a generating function for the 
tree form-factors in the theory, that is its expansion in powers of $a$'s reads
\beq
\label{3}
{\phi}^{ptb}(z, {\bar z}, \{k\}, \{a\})=\sum_{J_{d}}
<k_{j_{1}}, \ldots , k_{j_{d}}|{\phi}(z,{\bar z})|0>_{tree}
a_{j_{1}} \ldots a_{j_{d}}
\eeq
where the sum runs over all subsets $J_{d}$ of the set $1, \ldots , N$.
{\footnote{Since we work in the tree approximation we do not care about
separation of the positive- and negative-energy one-particle states. We just consider
all particles in the out-states while, more precisely,
those with negative energy should be considered as in-states.}} The nilpotency
of $a$'s means that the form-factors with identical one-particle states   
will not appear in the expansion Eq.(\ref{3}).

\section{Zero-curvature representation}

Zero-curvature representation for the Sin(h)-Gordon theory 
(see the book \cite{ZC} and references therein) can be taken in the 
form 
\begin{eqnarray}
\label{4}
A_{z}=-\frac{\beta}{4}{\sigma}_{1}{\partial}{\phi}+
\frac{\lambda}{2}{\sigma}_{3}cosh\frac{\beta{\phi}}{2}+
\frac{\lambda}{2}i{\sigma}_{2}sinh\frac{\beta{\phi}}{2}\nonumber\\
A_{\bar z}=\frac{\beta}{4}{\sigma}_{1}{\bar {\partial}}{\phi}-
\frac{1}{2 \lambda}{\sigma}_{3}cosh\frac{\beta{\phi}}{2}+
\frac{1}{2 \lambda}i{\sigma}_{2}sinh\frac{\beta{\phi}}{2}
\end{eqnarray}
where $\lambda$ is a non-homogeneous coordinate on an auxiliary $CP^{1}$ space,
the so-called spectral parameter, and ${\sigma}_{i}$ are Pauli matrixes.
The Sin(h)-Gordon equation (\ref{1}) is equivalent to
\beq
\label{5}
{\partial}A_{\bar z}-{\bar {\partial}}A_{z}+[A_{z},A_{\bar z}]=0
\eeq

The connection form Eq.(\ref{4}) is meromorphic on the auxiliary $CP^{1}$
space with simple poles at $\lambda=0$ and $\lambda=\infty$. Correspondingly,
the zero-curvature condition Eq.(\ref{5}) consists in fact of a number of 
equations - at different powers of $\lambda$ - most of which are automatically
resolved when the connection form is taken in the form Eq.(\ref{4}), independently
of the field ${\phi}(z,{\bar z})$. The only nontrivial equation arises at
${\lambda}^{0}$ and is equivalent to Eq.(\ref{1}). 

 For the following construction it is important that  a generic zero-curvature
connection obeying the ``reduction condition''
\beq
\label{6}
A(-{\lambda})={\sigma}_{1}A({\lambda}){\sigma}_{1}
\eeq
is equivalent to the connection Eq.(\ref{4}) modulo gauge transformations, 
choice of coordinates $z$, ${\bar z}$ and redefinition of the field
$\phi$ \cite{A.Mi}.  The gauge transformations are transformations 
\beq
\label{7}
A {\rightarrow} h^{-1}Ah+h^{-1}dh
\eeq
where $h$ is an $SL(2,C)$ matrix independent of $\lambda$ and commuting with the
reduction Eq.(\ref{6}), $d$ is external derivative. Notice that the field $\phi$ is 
a gauge invariant object.

The zero-curvature condition Eq.(\ref{5}) is (locally) solved as
\beq
\label{8}
A=g^{-1}dg
\eeq
where, unlike $h$ in Eq.(\ref{7}), $g$ is a nontrivial function of $\lambda$ subject to
the condition that the connection form $A(\lambda)$ has  simple poles
at $\lambda=0$ and $\lambda=\infty$ and also that  $A(\lambda)$ obeys
the reduction condition Eq.(\ref{6}) which for $g$ gives
\beq
\label{9}
g(-{\lambda})={\sigma}_{1}g({\lambda}){\sigma}_{1}
\eeq
The gauge transformation Eq.(\ref{7}) acts on $g(\lambda)$ as multiplication
by an independent of $\lambda$ matrix $h$ from the right.

Since ${\phi}^{ptb}$ is polynomial in $\{{\ej}\}_{j}$, the corresponding connection 
form $A^{ptb}$ is also polynomial. It is convenient to split off explicitly the zeroth
order part of it,
\beq
\label{10}
 A^{ptb}(\lambda, \{{\ej}\})=A^{(0)}(\lambda)+A'(\lambda, \{{\ej}\}),
\eeq
to define a derivative
\beq
\label{11}
{\nabla}^{(0)}=d+A^{(0)},
\eeq
to introduce $g^{(0)}(\lambda)$ such that
\beq
\label{12}
A^{(0)}={g^{(0)}}^{-1}dg^{(0)},
\eeq
and to define $g'(\lambda)$ according to
\begin{eqnarray}
\label{13}
A^{ptb}={g^{ptb}}^{-1}dg^{ptb},\nonumber\\
g^{ptb}(\lambda)=g^{(0)}(\lambda)g'(\lambda)
\end{eqnarray}
In terms of the introduced notations the local solution of the zero-curvature condition
rewrites as
\beq
\label{14}
A'={g'}^{-1}{\nabla}^{(0)}g'
\eeq
where the non-derivative term in ${\nabla}^{(0)}$ acts on $g'$ as commutator.

Like ${\phi}^{ptb}$, $g'$ can be sought for as polynomial in ${\ej}$'s. We stress
that, like in ${\phi}^{ptb}$, the space-time dependence of $g'$ comes only
via polynomials in the nilpotent variables ${\ej}$; that is why it was convenient
to split out  $g^{(0)}$ (which has different type of the space-time dependence).
First order terms $g'^{(1)}$ of the polynomial $g'$
are fixed by the ones in $A'$
which, in turn, are defined by the ones in ${\phi}^{ptb}$, Eq.(\ref{2}),
\beq
\label{15}
g'^{(1)}(\lambda)=\sum_{j=1}^{N}\frac{\beta}{4}{\ej}_{j}{\sigma}_{+}
\frac{{\lambda}+q_{j}}{{\lambda}+ik_{j}}
\frac{2ik_{j}}{ik_{j}-q_{j}}+
\frac{\beta}{4}{\ej}_{j}{\sigma}_{-}
\frac{{\lambda}-q_{j}}{{\lambda}-ik_{j}}
\frac{\beta}{4}{\ej}_{j}{\sigma}_{-},
\eeq
where ${\sigma}_{\pm}=\frac{1}{2}({\sigma}_{1}{\pm}i{\sigma}_{2})$.
{\footnote{Of course, the zeroth order term in the polynomial $g'$ is unit matrix,
$g'^{(0)}=1$.}}
The parameters $q_{j}$ in Eq.(\ref{15}) are present due to
the gauge freedom, Eq.(\ref{7}). To get the gauge condition used in Eq.(\ref{4})
one should put $q_{j}=-ik_{j}$, however there might be other convenient choice,
as we are anyway interested in the gauge invariant object, field $\phi$. 
 We remind that
that  index $j=1, \ldots , N$ numbers the one-particle states in the given set. 

These first order terms, $g'^{(1)}(\lambda)$ Eq.(\ref{15}),  possess simple poles
at $\lambda={\pm}ik_{j}$. From the condition of regularity of $A'$ Eq.(\ref{14})
anywhere except $\lambda=0$, $\lambda=\infty$ one can show that
the all the polynomial $g'(\lambda)$  can have only simple poles at the same
points, $\lambda={\pm}ik_{j}$, the higher order terms of the polynomial
being defined uniquely modulo gauge transformations.

To put the problem of finding $g'(\lambda)$ in a more universal form we
introduce an index ${\hat j}$ consisting of two indices,
\beq
\label{16}
{\hat j}=(j,s); j=1, \ldots, N; s={\pm},
\eeq
introduce notations ${\hej}_{\hat j}$,
\begin{eqnarray}
\label{17}
{\hej}_{j,+}=\frac{\beta}{4}\frac{2ik_{j}}{ik_{j}-q_{j}}
{\ej}_{j}{\sigma}_{+}\nonumber\\
{\hej}_{j,-}=\frac{\beta}{4}\frac{2ik_{j}}{ik_{j}-q_{j}}
{\ej}_{j}{\sigma}_{-},
\end{eqnarray}
and notations ${\lambda}_{\hat j}$, $q_{\hat j}$, where
\beq
\label{18}
{\lambda}_{j,{\pm}}={\mp}ik_{j}, q_{j,{\pm}}={\mp}q_{j}
\eeq
With these notations Eq.(\ref{15}) becomes
\beq
\label{19}
g'^{(1)}(\lambda)=\sum_{\hat j}{\hej}_{\hat j}
\frac{{\lambda}-q_{\hat j}}{{\lambda}-{\lambda}_{\hat j}}
\eeq
Introduce also $g'_{\hat j}$,
\beq
\label{20}
g'_{\hat j}(\lambda)=1+
{\hej}_{\hat j}\frac{{\lambda}-q_{\hat j}}{{\lambda}-{\lambda}_{\hat j}}
\eeq
Then the condition of regularity of the connection form $A'$ Eq.(\ref{14}) in
vicinity of point $\lambda={\lambda}_{\hat j}$ requires that
\beq
\label{21}
g'_{\hat j}(\lambda)^{-1}g'(\lambda)\; {\rm is\; regular\; at}\;
 \lambda={\lambda}_{\hat j}
\eeq

Thus, we are looking for $g'(\lambda)$, polynomial in ${\hej}_{\hat j}$ Eq.(\ref{17}),
first order term of the polynomial being as in Eq.(\ref{19}). $g'(\lambda)$
has simple poles at $\lambda={\lambda}_{\hat j}$ Eq.(\ref{18}) and obeys 
conditions Eq.(\ref{21}) at the poles. $g'(\lambda)$ must also obey the 
``reduction condition'' Eq.(\ref{9}). This problem for $g'(\lambda)$ has
a unique solution modulo the  gauge transformations.

\section{Solution}

The problem formulated in the last paragraph of section 2 is analogous to the ones
solved in the case of Yang-Mills theory in \cite{RS2} (that solution was later used
in the case of gravity in \cite{RS3} and in the case of Yang-Mills+gravity
in \cite{S1}). The new things are the reduction condition Eq.(\ref{9})
and the condition ${\hej}_{j,+}{\hej}_{j,-}=0$ (a trivial consequence of the nilpotency
condition, $a_{j}^{2}=0$), which are not crucial. We just give the solution,
\beq
\label{22}
g'(\lambda)=1+\sum_{d=1}^{N}\sum_{{\hat j_{1}}, \ldots , {\hat j_{d}}}
\frac{{\lambda}-q_{\hat j_{1}}}{{\lambda}-{\lambda}_{\hat j_{1}}}
\frac{{\lambda}_{\hat j_{1}}-q_{\hat j_{2}}}
{{\lambda}_{\hat j_{1}}-{\lambda}_{\hat j_{2}}}{\cdots}
\frac{{\lambda}_{\hat j_{d-1}}-q_{\hat j_{d}}}
{{\lambda}_{\hat j_{d-1}}-{\lambda}_{\hat j_{d}}}
{\hej}_{\hat j_{1}} {\ldots} {\hej}_{\hat j_{d}},
\eeq
all notations are introduced in Eqs.(\ref{16})-(\ref{18}). {\footnote{We point out a minor
subtlety concerning the solution Eq.(\ref{22}). At general values of the free parameters
$q_{j}$, $g'(\lambda)$ does not belong to $SL(2,C)$, only to $GL(2,C)$ instead.
However, one can show that ${\rm det} g'(\lambda)$ is, actually, independent 
of $\lambda$,
thus $g'(\lambda)$ is gauge equivalent to an $SL(2,C)$ matrix. Therefore this
subtlety doesn't actually matter since the field $\phi$ - which is the final aim -
is a gauge invariant object. On can also show that ${\rm det} g'(\lambda)=1$ when all
$q$'s are equal each other.}}
 The inverse for this matrix
reads
\beq
\label{23}
g'^{-1}(\lambda)=1-\sum_{d=1}^{N}\sum_{{\hat j_{1}}, \ldots , {\hat j_{d}}}
\frac{{\lambda}-q_{\hat j_{1}}}{{\lambda}-{\lambda}_{\hat j_{d}}}
\frac{{\lambda}_{\hat j_{2}}-q_{\hat j_{2}}}
{{\lambda}_{\hat j_{1}}-{\lambda}_{\hat j_{2}}}{\cdots}
\frac{{\lambda}_{\hat j_{d}}-q_{\hat j_{d}}}
{{\lambda}_{\hat j_{d-1}}-{\lambda}_{\hat j_{d}}}
{\hej}_{\hat j_{1}} {\ldots} {\hej}_{\hat j_{d}},
\eeq
With the Eqs.(\ref{22}),(\ref{23}) one can find the connection form $A'$ Eq.(\ref{14}),
from which one obtains, finally, ${\phi}^{ptb}$,
\beq
\label{24}
{\phi}^{ptb}=\sum_{d\;odd}\frac{2}{d}(\frac{\beta}{2})^{d-1}
\sum_{{j_{1}}, \ldots , {j_{d}}}
\frac{k_{j_{1}} {\ldots} k_{j_{d}}}
{(k_{j_{1}}+k_{j_{2}}) {\ldots} (k_{j_{d}}+k_{j_{1}})}
{\ej}_{j_{1}} {\ldots} {\ej}_{j_{d}}
\eeq
We remind that ${\ej}_{j}=a_{j}e^{ik_{j}z+i\frac{1}{k_{j}}{\bar z}}$
(due to the nilpotency, ${\ej}_{j}^{2}=0$, all
$j_{l}, l=1, {\ldots}, d$ must be different). ${\phi}^{ptb}$ is a generating function
for the tree form-factors in the sense of Eq.(\ref{3}).

{\bf Acknowledgments}\\
The work of A.R. was partially supported by the Grant 
No. RFBR-96-02-18046 and by the Grant No. 96-15-96455 for the support of 
scientific schools. Research of K.S. was  partially supported by  INTAS-96-482.


\begin{thebibliography}{40}
\bibitem{RS1}\ A.Rosly, K.Selivanov, preprint ITEP-TH-43-96,
 hep-th/9610070
\bibitem{FS}\ A.A.Slavnov, L.D.Faddeev, Introduction to the Theory of Quantum
 Gauge Fields,Nauka, Moscow, 1978
\bibitem{IZ}\ C.Itzykson, J.-B.Zuber, Quantum Field Theory, McGrow-hill, NY, 1980
\bibitem{RS2}\ A.Rosly, K.Selivanov, Phys.Lett.B 399 (1997) 135-140,
 hep-th/9611101
\bibitem{RS3}\ A.Rosly, K.Selivanov, preprint ITEP-TH-56-97,
 hep-th/9610196
\bibitem{S1}\ K.Selivanov, preprint ITEP-TH-59-97,
 hep-th/9710197
\bibitem{Ward}\ R.S.Ward, Phys.Lett.61A (1977) 81
\bibitem{Penrose}\ R.Penrose, Gen.Rel.Grav.,7 (1976) 31-52, 
 {\it ibid} 171-176
\bibitem{AHS}\ M.Atiyah, N.Hitchin, I.Singer, Proc.R.Soc.(London) A362 (1978) 425
\bibitem{S2}\ K.Selivanov, preprint ITEP-TH-64-97, \\
hep-th/9711111
\bibitem{PT}\ S.Parke, T.Taylor,  Phys.Rev.Lett. 56 (1986) 2459
\bibitem{BG}\ F.Berends, W.Giele, Nucl.Phys. B306 (1988) 759
\bibitem{Ba}\ W.Bardeen, Prog.Theor.Phys.Suppl, 123 (1996) 1
\bibitem{Se}\ K.Selivanov, preprint ITEP-21-96, hep-ph/9604206
\bibitem{KO}\ V.Korepin, T.Oota, J.Phys.A 29 (1996) 625, 
 hep-th/9608064
\bibitem{F.S1}\ F.Smirnov, Form-factors in completely integrable
 models of quantum field theory, Singapore, World Scientific, 1992
\bibitem{FMS}\ A.Fring, G.Mussardo, P.Simonetti, Nucl.Phys.B393 (1993) 413
\bibitem{F.S2}\ F.Smirnov, Comm.Math.Phys. 155 (1993) 459
\bibitem{Vergeles}\ S.Vergeles, V.Gryanik, Yad.Fiz.23 (1976) 1334 (in russian)
\bibitem{BFKL}\ L.N.Lipatov, Sov.J.Nucl.Phys. 23 (1976) 642\\
 V.S.Fadin, E.A.Kuraev, L.N.Lipatov, Phys.Lett.B 60 (1975) 50\\
 E.A.Kuraev, L.N.Lipatov, V.S.Fadin, Sov.Phys.JETP 44 (1976) 45,\\
 {\it ibid} 45 (1977) 199\\
  Ya.Ya.Balitsky, L.N.Lipatov, Sov.J.Nucl.Phys. 28 (1978) 822
\bibitem{ZC} \ L.D.Faddeev, L.A.Takhtajan, Hamiltonian methods in the theory 
of solitons, Moscow, Nauka, 1986 
\bibitem{A.Mi}\  A.Mikhailov, Physica D, 3 (1981) 73

\end{thebibliography}
\end{document}